\documentclass[12pt]{article}
\usepackage[pctex32]{graphics}
\begin{document}
\vspace{2cm}
\begin{center}
{\bf  \Large   Quantum Perfect-Fluid Kaluza-Klein Cosmology}
\vspace{1cm}

                      Wung-Hong Huang*  and  I-Chin Wang\\
                       Department of Physics\\
                       National Cheng Kung University\\
                       Tainan,70101,Taiwan\\

\end{center}
\vspace{2cm}

     The perfect fluid cosmology in the 1+d+D dimensional Kaluza-Klein spacetimes for an arbitrary barotropic equation of state $p = (\gamma -1) \rho$ is quantized by using the Schutz's variational formalism.    We make efforts in the mathematics to solve the problems in two cases.   In the first case of the stiff fluid $ \gamma = 2$ we exactly solve the Wheeler-DeWitt equation when the $d$ space is flat.   After the superposition of the solutions the wave-packet function is obtained exactly.   We analyze the Bohmian trajectories of the final-stage wave-packet functions and show that the scale functions of the flat $d$ spaces and the compact $D$ spaces will eventually evolve into the non-zero finite values.   In the second case of  $\gamma \approx 2$, we use the approximated wavefunction in the Wheeler-DeWitt equation to find the analytic forms of the final-stage wave-packet functions.   After analyzing the Bohmian trajectories we show that the flat $d$ spaces will be expanding forever while the scale function of the contracting $D$ spaces would not become zero within finite time.   Our investigations indicate that the quantum effect in the quantum perfect-fluid cosmology could prevent the extra compact $D$ spaces in the Kaluza-Klein theory from collapsing into a singularity or that the "crack-of-doom" singularity of the extra compact dimensions  is made to occur at $t=\infty$.

\vspace{2cm}
\begin{flushleft}
*E-mail:  whhwung@mail.ncku.edu.tw\\
\end{flushleft}

%%%%%%%%%%%%%%%%%%%%%%%

\newpage
\section{Introduction}

   It is known that over the distances ranging from $10^{-16}$ to $10^{25}$ cm our spacetime is four dimensional which is necessary for the existence of stable state, caused by long range force.  However, the theories of supergravity in 11 dimensions and superstring in 10 dimensions [1,2] indicate that the multidimensionality of space over the distance $r \ll 10^{-16} $ cm is necessary for the unification of all known types of interactions including the gravity.  The agreement between the multidimensional theory for the short distance and four-dimensional theory can be established through the mechanism of cosmologically dimensional reduction [3,4].   According to the mechanism the extra spaces will be contracting to a small value, at which it is assumed that the quantum gravity effects could support it from collapsing into a singularity.   

   Historically,  Sahdev [5] had shown a possible scenario in which the universe can pass from the multi-dimensional phase to the four dimensional Friedman-Robertson-Walker  universe, after the numerical integration.   The theory considered therein is the radiation-dominated with a perfect fluid $p=\rho/(d+D)$, in which $d+D$ is the dimensions of space.   The solution shows the "crack-of-doom" singularity in which the extra dimensions goes to zero at finite time. 

    In this paper we will  follow the references [6-10] to quantize the perfect-fluid cosmology in the 1+d+D dimensional Kaluza-Klein spacetimes by using the Schutz's variational formalism [11,12].  Our main motivation is to see whether the quantized perfect-fluid cosmology could prevent the extra compact spaces in the Kaluza-Klein theory from collapsing into a singularity. Our investigations show that for the stiff matter the scale functions are always non-zero and for some other matters the crack-of-doom singularity of the extra dimensions can be made to occur at $t=\infty$, so that it is not necessary a serious problem. 

    The quantum perfect fluid cosmology was first studied by Lapchinskii and Rubakov [6],  in which the Schutz's variables [11] permits us  to identify the matter field with time.   As the associated momentum appears linearly in the Lagrangian the restriction to minisuperspace has the advantage of allowing an explicit integration of the resulting Schr\"ondinger-like equation.   Along this line, the quantum perfect fluid cosmology for an arbitrary equation of state $p = n \rho $ has been investigated for the theory in the close, flat and open Friedmann-Robertson-Walker spacetimes [6-9].    In recent papers [10] Alvarenga et. cl. had also discussed the problem in the anisotropic Bianchi I spacetime.    

   This paper is organized as follows.  In section 2 we briefly describe the method of constructing perfect-fluid action and its quantization [6,11,12].  In section 3, we quantize the gravitational theory in the 1+d+D dimensional spacetime with the perfect fluid of arbitrary barotropic equation of state $ p = (\gamma -1) \rho$ and obtain the associated Wheeler-Dewitt equation.  In section 4,  the exact solutions of the Wheeler-Dewitt equation are found when the $d$ space is flat for the case with stiff fluid $p=\rho$, which can effectively describe the matters with quantum effects in the early universe [13].   After the superposition of the solutions  we can, due to the mathematical difficulty, obtain the analytic forms of the wave-packet functions only for the universe in the final stage.    The time evolution of the scale factors are then determined by computing the Bohmian trajectories of the ontological interpretation as that in [7-10].   Our results show that the flat $d$ spaces and the compact $D$ spaces will eventually evolve into finite scale functions.   The scale functions do not become zero within finite time.   However, the mechanism of cosmologically dimensional reduction does not shown in the stiff-matter spacetime.   In section 5 we turn to the cases of  $\gamma \approx 2$.    We use the approximated wavefunction in the Wheeler-DeWitt equation to find the analytic forms of the wave-packet functions in the final stage of the universe.   The time evolution of the scale factors are also determined by computing the Bohmian trajectories.   The results show that the flat $d$ spaces will be expanding forever while the scale function of the $D$ spaces is contracting and does not become zero within finite time.   Our investigations indicate that the quantum effect in the quantum perfect-fluid cosmology could prevent the extra compact $D$ spaces in the Kaluza-Klein theory from collapsing into a singularity or that the "crack-of-doom" singularity in which the extra dimensions goes to zero at finite time in the classical model [5] is now made to occur at $t=\infty$.   Last section is devoted to a short conclusion.

%%%%%%%%%%%%%%%
\section{Action of Perfect Fluid and Quantization}
\subsection{Action of Perfect Fluid}
Perfect fluid can be described by various thermodynamical variables. In the notation of Misner, Thorne, and Wheeler [14], they are
$$ n = {\hbox{particle number density}},$$
$$  \rho = {\hbox{energy density}} ,~~~~~~~~~~~~$$
$$ p = {\hbox{pressure}} ,~~~~~~~~~~~~~~~~~~~~$$
 $$S = {\hbox{entropy per particle}} \, .~~~~~~$$
These variables are spacetime scalar fields whose values represent measurements made in the rest frame of the fluid. 

   The fluid motion is characterized by its unit four-velocity vector field $U_{\mu}$.  In the Schutz's formalism [11] the four-velocity is expressed in terms of the five potentials $\epsilon$,  $\zeta$, $\phi$, $\theta$ and $S$:
$$  U_\mu = \frac{1}{h}\, (\epsilon_{,\mu} + \zeta\phi_{,\mu} +
\theta S_{,\mu}), \eqno{(2.1)}$$
where $h$ is the specific enthalpy. The potentials $\zeta$ and $\phi$ are connected with rotation and are absent, for example, in the  FRW's spacetime. The variables $\epsilon$ and $\theta$ are the auxiliary fields which have no clear physical meaning [11,12].   The four velocity is subject to the condition 
$$  U^\mu U_\mu = -1, \eqno{(2.2)}$$
which then implies that
$$h^2 = - g^{\mu\nu}\, (\epsilon_{,\mu} + \zeta\phi_{,\mu} +
\theta S_{,\mu})\, (\epsilon_{,\nu} + \zeta\phi_{,\nu} +
\theta S_{,\nu}). \eqno{(2.3)}$$
From (2.1) and (2.3) we see that
$$\delta (h^2) = 2\, h \,\delta h = - \delta g^{\mu\nu} \, h \, U^\mu \, h \, U^\nu. \eqno{(2.4)}$$
Thus 
$$\delta h = - {h\over 2} U^\mu \, U^\nu \, \delta g^{\mu\nu} \, = -{1\over 2}{\rho + p\over n} U^\mu \, U^\nu \, \delta g^{\mu\nu}, \eqno{(2.5)}$$
in which we have used a thermodynamical relation [14]
$$h = {1\over n} (\rho + p), \eqno{(2.6)}$$

   Finally, using the thermodynamical relation ($\partial p/ \partial h = n$), equation (2.5), and the property that the pressure of the prefect fluid is the function of enthalpy ($p= p(h)$) we then have the relation 
$$\delta p = {\partial p\over \partial h} \, \delta h = n\,  \delta h= - {1\over 2}(\rho + p)\, U^\mu \, U^\nu \, \delta g^{\mu\nu}. \eqno{(2.7)}$$
Therefore
$$\delta(p \sqrt{-g}) = (\delta \,p) \,\sqrt{-g} +  p \,(\delta \,\sqrt{-g}) = - {1\over 2}{\Large [}(\rho + p)\, U_\mu \, U_\nu \,+ p \, g_{\mu\nu}{\Large ]}\,\delta \,g^{\mu\nu} = - {1\over 2}\, T_{\mu\nu}\, \delta \,g^{\mu\nu}. \eqno{(2.8)}$$
This shows that the Lagrangian of perfect fluid may be expressed as 
$$L = p \sqrt{-g}, \eqno{(2.9)}$$
because its variation with respect to the metric $g_{\mu\nu}$ could produce the well-known stress energy tensor of the perfect fluid $T_{\mu\nu}$.

\subsection{Quantization of Perfect Fluid}
  To quantize the perfect-fluid action we will consider the homogeneous spacetime without rotation.   In this case the potential  $\zeta$ and $\phi$ are zero [11] and thus (2.3) becomes
$$h^2  = -g^{00}\, (\dot \epsilon +  \theta \dot S)^2 .\eqno{(2.10)}$$
For a perfect fluid in which the pressure $p$ is linked to the energy density by the equation of state $p = (\gamma - 1)\, \rho$ we have a solution [14]
$$ \rho ={h\over \gamma} \, e^{-S}.  \eqno{(2.11)}$$
Thus
$$p = (\gamma - 1) \rho = (\gamma - 1)\, \left({h\over \gamma}\right)^{\gamma\over \gamma -1} \, e^{-S} = (\gamma - 1)\, (\gamma) ^{\gamma\over \gamma -1} \left(\dot \epsilon +  \theta \dot S\right) ^{\gamma\over \gamma -1} \, e^{-S}.          \eqno{(2.12)}$$
To obtain this result we have adopted the commoving frame with $g^{00} = -1$. 

 Using the above relation the perfect-fluid Hamiltonian becomes 
$$H_{PF}[\epsilon, \, \Pi_{\epsilon}, \, S, \,\Pi_S] = \dot \epsilon \, \Pi_{\epsilon}+  \dot S \, \Pi_S - L  = \sqrt{-g}\, \left(\Pi_{\epsilon}\right)^\gamma \,e^{-(1-\gamma)\,S}, \eqno{(2.13)}$$
in which 
$$\Pi_{\epsilon} = {\partial L\over \partial \epsilon}, \eqno{(2.14a)}$$
$$\Pi_{S} = {\partial L\over \partial S}, \eqno{(2.14b)}$$
In a remarkable paper Lapchinskii and Rubakov [6] have found that the above perfect-fluid Hamiltonian  may be recast in a more suitable form through the canonical transformation  $(\epsilon, \, \Pi_{\epsilon}, \, S, \,\Pi_S) \rightarrow (T, P_T,\bar\epsilon,  \bar P_\epsilon)$ by the following relation
$$T = P_S e^{-S}P_\epsilon^{-(\gamma + 1)} \quad , \quad 
P_T = P_\epsilon^{\gamma + 1}e^S \quad , \quad
\bar\epsilon = \epsilon - (\gamma + 1)\frac{P_S}{P_\epsilon} \quad ,
\quad \bar P_\epsilon = P_\epsilon \quad.  \eqno{(2.15)}$$
After some calculations they found that the perfect-fluid Hamiltonian becomes an elegant expression
$$ H_{PF}[T, P_T] = \sqrt{-g} \, P_T ,\eqno{(2.16)}$$
in which $T$ is a new canonical coordinate and the associated canonical momentum $P_T$ is the only remaining variable in the new Hamiltonian $H_{PF}[T, P_T]$.  Equation (2.16) is linear in $T$ and as argued by Lapchinskii and Rubakov [6]  $T$ may be identified as the cosmic time.

%%%%%%%%%%%%%%%%%%%%%%%%
\section{Wheeler-DeWitt Equation of Perfect-Fluid Kaluza-Klein Theory}
The action of $1+d+D$-dimensional gravity coupled to a perfect fluid in the Schutz's formalism is [6-10]
$$ S = \int_Md^{1+d+D}x\sqrt{-g}(R + p) + 2\int_{\partial M}d^{1+d+D}x\sqrt{h}h_{ab}K^{ab}, \eqno{(3.1)}$$
where $K^{ab}$ is the extrinsic curvature and $h_{ab}$ is the induced
metric over the $d+D$-dimensional spatial hypersurface.   

    We consider the theory describing an $1+d+D$-dimensional anisotropic spacetime with the following metric 
$$ ds^2 = -N(t)^2dt^2 + e^{2\alpha (t)}g_{\tilde n \tilde m} + e^{2\beta(t)}g_{\tilde N \tilde M}, \eqno{(3.2)}$$
where $\tilde n,\tilde m = 1,2,...,d$;  $\tilde N,\tilde M = d+1, d+2,...,d+D$.  $g_{\tilde n\tilde m}$ and $g_{\tilde N\tilde M}$ are the metrics  for the maximally symmetric  d and D-dimensional spaces respectively and $e^{\alpha (t)}$, $e^{\beta(t)}$ the corresponding time-dependent cosmological scale functions.   In this expression $N(t)$ is the lapse function.

   The curvature for the metric (3.2) is
$$  R(t) = (d-d^2)\dot \alpha^2 + (D-D^2) \dot \beta^2 - 2dD \dot\alpha \dot\beta +  2 k_d ~d ~e^{-2\alpha} +2 k_D D e^{-2\beta}, \eqno{(3.3)}$$
in which $k_{d,D}= 1,0$ or ,$-1$.   After the definition
$$ \alpha = \sqrt{D-1\over d+D-1}~ X, \hspace{4.6cm} \eqno{(3.4a)}$$
$$ \beta  = - \sqrt{d\over (D-1)(d+D-1)}~ X + {1\over \sqrt{D^2-D}}~ Y, \eqno{(3.4b)}$$
the curvature becomes  
$$  R(t) =  \dot X^2 -  \dot Y^2 +  2 k_d ~d ~e^{-2  \sqrt{D-1\over d+D-1}~ X} +2 k_D D e^{-2 \sqrt{d\over (D-1)(d+D-1)}~ X + 2 {1\over \sqrt{D^2-D}}~ Y}. \eqno{(3.5)}$$
Using the above relation the gravitational Hamiltonian in the Kaluza-Klein spacetime (3.2) has a simple form
$$ H_G = {1\over 4} ~ e^{-\Delta} \left\{ P_X^2 - P_Y^2 - 8Dk_D e^{2 \sqrt{D-1\over D}~ Y} - \right.$$
$$\hspace{6cm}\left. 8d~ k_d~ e^{-2\left(\sqrt{d\over (D-1)(d+D-1)}+\sqrt{D-1\over d(d+D-1)}\right)~ X + 2 \sqrt{D\over D-1}~ Y}\right\} , \eqno{(3.6)}$$
in which the canonical momentums are defined by 
$$P_X \equiv  {\partial L\over \partial X } = 2 e^{ \Delta} \dot X , \hspace{6.4cm}\eqno{(3.7)}$$
$$P_Y \equiv {\partial L\over \partial Y } = - 2 e^{ \Delta} \dot Y , \hspace{6cm}\eqno{(3.8)}$$
and we have defined a new variable  
$$ \Delta \equiv - \sqrt{d\over (D-1)(d+D-1)}~ X + \sqrt{D\over D-1}~ Y.\eqno{(3.9)}$$

   Using the method described in section 2 the Hamiltonian of perfect fluid  has an elegant expression
$$ H_{PF} = e^{ (2 - \gamma) \left(- \sqrt{d\over (D-1)(d+D-1)}~ X + \sqrt{D\over D-1}Y\right)}~ P_T ,\eqno{(3.10)}$$
in which the momentum $P_T$ is the only remaining variable of the perfect fluid.  It is linear and may be identified as the {\it cosmic time} [6].  In this paper we will analyze the {\it time} evolutions of the quantum perfect fluid cosmology in the 1+d+D dimensional spacetime.

   Now, substituting the quantization relations
$$P_X \rightarrow -i \,{\partial \over \partial X} \, , \eqno{(3.11)}$$
$$P_Y \rightarrow -i \,{\partial \over \partial Y}\, , \eqno{(3.12)}$$
$$P_T~ \rightarrow ~ i \,{\partial \over \partial T}\, ,~ \eqno{(3.13)}$$
into (3.6) and (3.10) the Wheeler-DeWitt equation $\left(H_G + H_{PF}\right) \, \Psi(T,X,Y) = 0$ becomes

$$ \left[-{\partial^2 \over \partial X^2} + {\partial^2 \over \partial Y^2}- 8 D  k_De^{2 \sqrt{D-1\over D}~ Y}-  8d~ k_d ~ e^{-2\left(\sqrt{d\over (D-1)(d+D-1)}+\sqrt{D-1\over d(d+D-1)}\right)~ X + 2 \sqrt{D\over D-1}~ Y}\right.$$

$$\left. i ~4~e^{ (2 - \gamma) \left(- \sqrt{d\over (D-1)(d+D-1)}~ X + \sqrt{D\over D-1}Y\right)}~ {\partial  \over \partial T}\right] \Psi(T,X,Y) = 0.\eqno{(3.14)}$$

   Using the above formula we will in the next section study the case of stiff fluid $p=\rho$ and then in the section 5 turn to the cases of  $\gamma \approx 2$.    We will consider only the case of  $k_d=0$ with $k_D=1$, otherwise the wavefunction could not be obtained.   We will see that for these cases the contracting scale function does not become zero within finite time.   This indicate that the quantum effect in the quantum perfect-fluid cosmology could prevent the extra compact $D$ spaces in the Kaluza-Klein theory from collapsing into a singularity.  
%%%%%%%%%%%%%%%

\section{Quantum Perfect-Fluid Kaluza-Klein Cosmology with Stiff  Fluid $p=\rho$} 
It is known that the spacetimes in the early universe may be regarded as that full of stiff fluid $p = \rho$ [13].   In this case $\gamma = 2$  and the wavefunction can be chosen as  
$$\Psi_{E,\pm\nu}(T,X,Y) = e^{- iET}~ e^{\pm i \nu X} ~ \Phi_{E,\pm\nu} (Y).   \eqno{(4.1)}$$
in which, using (3.14),  the function $\Phi_{E,\pm\nu} (Y)$ shall satisfy the  equation
$$ {d^2\over dY^2}\Phi_{E,\pm\nu}(Y) + \left[(\nu^2 + 4 E) - 8 D e^{2 \sqrt{D\over D-1}~ Y}\right]\Phi_{E,\pm\nu} (Y) = 0.  \eqno{(4.2)}$$
The above equation has the Bessel function $Z_{i z}$ as its exact solution [15]
$$\Phi_{E,\pm\nu} (Y) = Z_{i {\sqrt{D(\nu^2 + 4 E)/(D-1)}}}\left(w\right) ,  \eqno{(4.3)}$$
with 
$$ w\equiv \sqrt{8D/( D-1)}~ e^{\sqrt{D-1\over D}Y}.  \eqno{(4.4)}$$
Now following the Misner [16] we can linearly combine the functions $\Psi_{E,\pm\nu}(T,X,Y)$ to choose the wavefunction 
$$\tilde\Psi_{E,\nu}(T,X,Y) = e^{- iET}~ sin(\nu X) ~ K_{i {\sqrt{D(\nu^2 + 4 E)/(D-1)}}}\left(w\right) ,   \eqno{(4.5)}$$
in which $K_{i z}$ is the modified Bessel function.  The above choice ensure that the wavefunction is regular everywhere.

    Note that the  quantum  Kantowski-Sachs cosmology, which corresponding to our case of  $d=1$ and $D=2$, had been discussed by Misner.   However, the model considered in [16] does not include the quantum perfect fluid and thus lacks the corresponding cosmic time $T$.

    We now have to find the wave-package function $\Psi(T, X, Y)$ by superpositing the mode wavefunction $\tilde\Psi_{E,\nu}(T,X,Y)$
$$\Psi(T, X, Y) = \int_0^\infty d\nu \int_0^\infty dE ~ e^{- iET}~ sin(\nu X) ~ K_{i {\sqrt{D(\nu^2 + 4 E)/ (D-1)}}} \left(w\right) , \eqno{(4.6)}$$  
\\
as  these done in [16].   To proceed, let us first use the new variables $r$ and $\theta$ defined by 
$$ \nu = \sqrt{(D-1)\over D}~ r~cos(\theta), ~~~~~~ E = {(D-1)\over 4D}~ r^2~sin(\theta)^2 \eqno{(4.7)}$$
to express (4.6) as 
$$\Psi(T, X, Y) = \int_0^\infty dr \int_0^\pi d\theta ~rcos(\theta) ~ e^{- i{(D-1)\over 4D} r^2sin(\theta)^2T}~ sin\left(\sqrt{(D-1)\over D}~ rXcos(\theta)\right) \times~$$
$$ K_{i r} \left(\sqrt{8D/( D-1)}~ e^{\sqrt{D-1\over D}Y}\right) . \eqno{(4.8)}$$  
However, as the integrals over the order $z$ in the modified Bessel function $K_{iz}$ is too difficult we will consider only the later-stage universe.    

   It is seen that in the later stage the exponential term in (4.8) will become a quickly oscillating function, as $T \rightarrow \infty$.   Therefore the main contribution of the integration shall be those from the small value of $r$.   Due to this property we can first substituting the expansion
$$K_{ir}(w)= K_{0}(w)+ ir K_{0}^{(1,0)}(w) - {r^2\over 2} K_{0}^{(2,0)}+ ...,\eqno{(4.9)}$$
in which 
$$K_{0}^{(k,0)}(w) \equiv {d^k\over dz^k}K_{z}(w)|_{z=0}, \eqno{(4.10)}$$  
into (4.8) to obtain an approximation
$$\Psi(T, X, Y) = \int_0^\infty dr \int_0^\pi d\theta ~rcos(\theta) ~ e^{- i{(D-1)\over 4D} r^2sin(\theta)^2T}~ sin\left(\sqrt{(D-1)\over D}~ rXcos(\theta)\right) \times~$$
$$ \left(K_{0}(w)  - {r^2\over 2}K_{0}^{(2,0)}(w) + ...\right). \eqno{(4.11)}$$
\\
Note that because  $K_{z}=K_{-z}$ [14] , $K_{0}^{(k,0)}(w)= 0$ if $k$ is odd.  

  Next, using the following two integration formulas [15]

$$\int_0^\infty dr ~r~ sin(a r^2)~ sin(2br) = {b\over 2a}\sqrt{\pi\over 2a}\left(cos\left({b^2\over a}\right)+ sin\left({b^2\over a}\right) \right), \eqno{(4.12a)}$$
$$\int_0^\infty dr ~r~ cos(a r^2)~ sin(2br) = {b\over 2a}\sqrt{\pi\over 2a}\left(-cos\left({b^2\over a}\right)+ sin\left({b^2\over a}\right) \right), \eqno{(4.12b)}$$
\\
Eq.(4.10) becomes 
$$\Psi(T, X, Y) \sim  \int_0^1ds ~ e^{ i{X^2 s^2\over T (1-s^2)}}~  {1\over \sqrt{1-s^2}}\left[{Xs\over 2}\left({(D-1)\over D}{T(1-s^2)\over 4} \right)^{-3\over 2} ~ K_0(w) - \right.$$

$$ \left.\left({i3Xs\over 4}\left({T(D-1)(1-s^2)\over 4D} \right)^{-5\over 2}- \left({D-1\over D}\right) {X^3s^3\over 8}\left({T(D-1)(1-s^2)\over 4D} \right)^{-7\over 2} \right)  K_0^{(2,0)}(w) +... \right],\eqno{(4.13)}$$
\\
in which $s \equiv cos(\theta)$.   Note that in the above (and hereafter) we will always neglect the irrelevant constant before a wavefunction.

   Finally, the integration over the variable $s$ can be performed through the following calculations

$$ \int_0^1ds ~ {2s\over (1-s^2)^2} ~e^{ i{X^2 s^2\over T (1-s^2)}}~ =  e^{- i{X^2 \over T }}~ \int_0^1ds ~ {2s\over (1-s^2)^2} ~e^{ i{X^2 \over T (1-s^2)}}~= e^{- i{X^2 \over T }}~\int_1^\infty du ~ 2u ~e^{ i{X^2 u^2 \over T}}$$

$$ = e^{- i{X^2 \over T }}~ \int_1^\infty dt ~  ~e^{ i{X^2 t \over T}}~ = e^{- i{X^2 \over T }}\left(~\int_0^\infty dt ~  ~e^{ i{X^2 t \over T}} - \int_0^1 dt ~e^{ i{X^2 t \over T}}\right) $$

$$ = e^{- i{X^2 \over T }}~ \left\{{1\over2}\,\,\delta\left({X^2 \over T}\right) - {T\over X^2 }\left[sin\left({X^2\over T}\right)-i \left(cos\left({X^2 \over T}\right)-1\right)\right]\right\}.  \eqno{(4.14)}$$
\\
Therefore we have the final result

$$ \Psi(T, X, Y) \sim  {1\over X \sqrt T }~ e^{- i{X^2 \over T }} K_{0}\left(\sqrt{8D\over( D-1)}~ e^{\sqrt{D-1\over D}Y}\right) \left[sin\left({X^2\over T}\right)-i \left(cos\left({X^2 \over T}\right) -1\right)\right]+ ...,  \eqno{(4.15)}$$
\\
in which the neglected terms are those proportional to  $ K_0^{(2,0)}(w)$.   These neglected terms can be evaluated in a similar way and the results are the order of ~$ T^{-1}$~ compared to the terms proportional to $ K_0(w)$, thus it is negligible.   

   To find the cosmological evolution from the equation (4.15) we shall calculate the expectation value of the scale factor, in the spirit of the many-worlds interpretation,  or analyze the Bohmian trajectories, on the ontological interpretation, to find the space-time dynamical behaviors [9].   However, it was argued [10] that there is inequivalence between these two interpretations due to the hyperbolic structure of the "Schr\"odinger-like" equation.   In this paper we will analyze the Bohmian trajectories of the final-stage wave-packet functions.
  
   In the ontological interpretation the wave function (4.15) is first written as 
$$ \Psi(T, X, Y) \sim  Re^{iS},  \eqno{(4.16)}$$
in which 
$$S= {- {X^2 \over T }} - tan^{-1}\left (cos\left({X^2 \over T} \right) -1 \over sin \left({X^2 \over T} \right)\right) = {- {3X^2 \over 2T }}. \eqno{(4.17)}$$
\\
the Bohmian trajectories are then determined by
$$P_X = {\partial S\over \partial X}, \eqno{(4.18)}$$
$$P_Y = {\partial S\over \partial Y}. \eqno{(4.19)}$$
Substituting Eqs.(3.7), (3.8) and (4.17) into Eqs.(4.18) and (4.19) we obtain the following equations

$$ 2 e^{ - \sqrt{d\over (D-1)(d+D-1)}~ X + \sqrt{D\over D-1}~ Y} \dot X  = - {3X \over T }\eqno{(4.20)}$$
$$ 2 e^{- \sqrt{d\over (D-1)(d+D-1)}~ X + \sqrt{D\over D-1}~ Y} \dot Y = 0, \hspace{0.6cm}\eqno{(4.21)}$$
Above equations tell us that $|X(T)|$ is a decreasing function and $Y(T)=Y(0)$.   For small values of $X$ Eq.(4.20) has the solution
$$ X(T)=X(0)\left(T_0\over T\right)^{c},\eqno{(4.22)}$$
in which $c$ is a positive number.  We thus see that the scalar function $X$ needs infinite time to arrive its minimum value.

   The results of this section tell us that the final stage of the quantum stiff fluid cosmology in the 1+d+D dimensional Kaluza-Klein spacetimes will has finite scale functions.   Thus the quantum effect in the quantum perfect-fluid cosmology could prevent the spaces from collapsing into a singularity.   Note that the classical stiff fluid cosmology is oscillating and it has zero scale functions at which the space is singular. 

  However, as the mechanism of cosmologically dimensional reduction does not shown in the stiff-matter spacetime we will therefore turn to the case with $\gamma  \approx 2$ in the next section.   

%%%%%%%%%%%%%%%%%

\section{Quantum Perfect-Fluid Kaluza-Klein Cosmology with $p \approx \rho$}
 When $p \approx \rho$, i.e. $\gamma \approx 2$, the Wheeler-DeWitt equation  (3.14) becomes

$$ \left\{-{\partial^2 \over \partial X^2} + {\partial^2 \over \partial Y^2}- 8 D e^{2 \sqrt{D\over D-1}~ Y}+ i ~4~\left[1 + (2 - \gamma) \left(- \sqrt{d\over (D-1)(d+D-1)}~ X +\right. \right. \right. $$
$$ \hspace{8cm}\left.\left.\left.\sqrt{D\over D-1}Y\right)\right]~ {\partial  \over \partial T}\right\} \Psi(T,X,Y) \approx  0.  \eqno{(5.1)}$$
\\
Above equation can be solved by choosing the wavefunction 
$$\Psi(T,X,Y) \approx  e^{-iET} \left[f_0(X) + f_n(X)\right] \left[g_0(Y) + g_n(Y)\right],  \eqno{(5.2)}$$
in which the functions $f_i(X)$ and $g_i(Y)$ satisfy the equations

$$ \left[-{\partial^2 \over \partial X^2} -~4E (2 - \gamma) \sqrt{d\over (D-1)(d+D-1)}~ X  -\nu^2\right] \left[f_0(X) + f_n(X)\right] = 0, \eqno{(5.3a)}$$
$$ \left[{\partial^2 \over \partial Y^2}- 8 D e^{2 \sqrt{D\over D-1}~ Y}-4E~ (2 - \gamma)  \sqrt{D\over D-1}Y + \nu^2 +4E \right] \left[g_0(Y) + g_n(Y)\right] = 0,\eqno{(5.3b)}$$
\\
where the zero-order solutions $f_0(X) = e^{\pm i \nu X}$ and $g_0(Y)=K_{iz}(w)$ are those obtained in the (4.5).

   In this section we will show that the small correction in the wavefunction coming from $f_n(X)$ will  be growing during time evolution and it will render the scale function $X(T)$ to become a time-increasing function, in contrast to that in the stiff matter system.     Therefore  the $1+d+D$ dimensional Kaluza-Klein spacetime with $\gamma \approx 2$ will have the expanding $d$ spaces and contracting compact $D$ spaces.    This thus realize the mechanism of cosmologically dimensional reduction [3-4].    Furthermore, the quantum version considered in this section also shows that the scale function of the contracting $D$ spaces would not become zero within finite time.   Thus the "crack-of-doom" singularity of the extra compact dimensions  is made to occur at $t=\infty$.

   To begin with we know that to the leading order, when $|\gamma -2| \ll 1$ the eq. (5.3a) becomes 

$$ \left[-{\partial^2 \over \partial X^2}  -\nu^2\right]  f_n(X) = 4E (2- \gamma) \sqrt{d\over (D-1)(d+D-1)}~ X ~ sin(\nu X), \eqno{(5.4)}$$
\\
which has the solution
$$ f_n(X) = (2- \gamma) E X^2~\sqrt{d\over (D-1)(d+D-1)} {cos(\nu X)\over \nu }+ ..., \eqno{(5.5)}$$
\\
in which we have neglected the terms which become smaller if $T \gg 1$.  Now we will use the function $f_n(X)$ and $g_0(Y)$ to evaluate the correction to the wave-packet function. 

    Following the prescription in section 4  we define the new variables $r$ and $\theta$ as that in (4.7).   Then, after the integrations over the variable $r$ we have the relation 

$$\Psi_{correct}(T, X, Y) = \int_0^\infty dr \int_0^\pi d\theta ~rcos(\theta) ~ e^{- i{(D-1)\over 4D} r^2sin(\theta)^2T}~(2- \gamma) E X^2 {cos(\nu X)\over \nu }\times~\hspace{2cm}$$
$$\hspace{5cm} K_{i r} \left(\sqrt{8D/( D-1)}~ e^{\sqrt{D-1\over D}Y}\right) $$  

$$= (2- \gamma) X^2 K_{0}\left(\sqrt{8D\over( D-1)}~ e^{\sqrt{D-1\over D}Y}\right) \int_0^\infty dr~ r\int_0^1 ds~ \times\hspace{3cm}$$
$$\hspace{5cm}e^{- i{(D-1)\over 4D} r^2T(1-s^2)}~ {1-s^2\over s}cos\left(\sqrt{(D-1/ D)}~ rXs)\right) + ...$$

$$= (2- \gamma) X^2 K_{0}\left(\sqrt{8D\over( D-1)}~ e^{\sqrt{D-1\over D}Y}\right) e^{- i{X^2\over T}}~~\int_0^1ds ~ e^{ i{X^2\over T (1-s^2)}}~  {1\over 1-s^2} + ... , \eqno{(5.6)}$$
\\
in which $s \equiv cos(\theta)$.   Note that the neglected terms, which are those coming from $K_0^{(2,0)}$, become smaller if $ T\gg 1$.   We have also neglected the irrelevant constant before the wavefunction. 

   The integration over the variable $s$ can be performed through the following calculations

$$ \int_0^1ds ~ {2\over (1-s^2)} ~e^{ i{X^2 \over T (1-s^2)}}~ = \int_1^\infty du ~ {1\over  \sqrt{u^2-1}} ~e^{ i{X^2 u^2 \over T}} = ~ \int_1^\infty dt ~{1\over \sqrt{t}\sqrt{t-1}}  ~e^{ i{X^2 t \over T}}~ \hspace{2cm}$$

$$ = e^{ i{X^2 \over T}} \int_0^\infty dz ~{1\over \sqrt{z}\sqrt{z+1}}  ~e^{ i{X^2 z \over T}}= 2 ~ e^{ i{X^2 \over T}} \int_0^\infty dv ~{1\over\sqrt{1+v^2}}  ~e^{ i{X^2 v^2 \over T}} \hspace{3.5cm}$$

$$ = {\pi \over2}e^{ i{X^2 \over T}}\left[ J_0\left(X^2\over2T\right)\left(sin\left(X^2\over2T\right) + i cos\left(X^2\over2T\right)\right)+N_0 \left(-cos\left(X^2\over2T\right) + i sin\left(X^2\over2T\right)\right)\right]$$

$$= {\pi \over2}e^{ i{X^2 \over T}}\left[\left({X^2\over2T} + i\right) +  {2\over \pi} \left(-1 + i {X^2\over2T}\right)~ln\left(X^2\over4T\right) \right] + ...,\hspace{4cm}\eqno{(5.7)}$$
\\
in which $J_0$ and $N_0$ are the Bessel functions  and the neglected terms are small at latter stage when $ T \gg 1$.

    The above correction is that coming from the function $f_n(X)  g_0(Y)$.    When comparing it with the zero-order term,  i.e.  the function $f_0(X)  g_0(Y)$ which had been evaluated in section 4, we see that the factor ${1/ \sqrt T}$ in eq.(4.15) shows a property that the zero-order term will be small if $T\gg 1$.   In a same way,  it can also be seen that the correction coming from the function $f_0(X)  g_n(Y)$ will also have a factor ${1/ \sqrt T}$ and thus this correction also becomes small if $T\gg 1$.

   Therefore we have the final result
$$ \Psi(T, X, Y) = \Psi_0(T, X, Y) + \Psi_{correct}(T, X, Y) \hspace{8cm}$$
$$\approx  (2 - \gamma) X^2 ~K_{0}\left(\sqrt{8D\over( D-1)}~ e^{\sqrt{D-1\over D}Y}\right) \left[\left({X^2\over2T} + i\right) +  {2\over \pi}\left(-1 + i {X^2\over2T}\right)~ln\left(X^2\over4T\right)\right],  \eqno{(5.8)}$$
\\
in which $\Psi_0$ is the zero-order term that calculated in (4.15), which is small comparing to the $\Psi_{correct}$ when $T\gg 1$. 
   
   We can now find the cosmological evolution by analyzing the Bohmian trajectories of the final-stage wave-packet functions.    Expressing the wave function (5.8) as the expression (4.16) we find that 
$$S = tan^{-1}\left(1+ {2\over\pi}{X^2\over 2T}ln\left({X^2\over 2T}\right)\over  {X^2\over 2T}-{2\over\pi}ln\left({X^2\over 2T}\right) \right) .\eqno{(5.9)}$$
In the later stage we see that 
$$S \approx {\pi\over 2} + tan^{-1}\left({2\over\pi}ln\left({X^2\over 2T}\right) \right), ~~~~~~~ T\gg 1  .\eqno{(5.10)}$$
The Bohmian trajectories determined by Eqs.(4.18) and (4.19) now become

$$ 2 e^{ - \sqrt{d\over (D-1)(d+D-1)}~ X + \sqrt{D\over D-1}~ Y} \dot X  = {\pi \over X \left[ln(X^2/2T)\right]^2 } ,\eqno{(5.11)}$$
$$ 2 e^{- \sqrt{d\over (D-1)(d+D-1)}~ X + \sqrt{D\over D-1}~ Y} \dot Y =  0, \hspace{1cm}\eqno{(5.12)}$$
\\
if $T\gg 1$.  Equation (5.12) implies that $Y(T) = Y(0)$ and (5.11) tells us that $|X(T)|$ is an increasing function.   Therefore we have shown that the flat $d$ spaces will be expanding forever while the scale function of the contracting $D$ spaces would not become zero within finite time.   Our investigations indicate that the quantum effect in the quantum perfect-fluid cosmology could make the "crack-of-doom" singularity, in which the extra dimensions goes to zero at finite time in the classical model [5], to occur at $t=\infty$.

%%%%%%%%%%%%%%%%%

\section{Conclusion}

The physics with extra spatial dimensions is known to play an important role in the unification of all known types of interactions including the gravity.    The existence of the extra dimensions is open to experimental investigation either now or in the foreseeable future. For the high-dimensional theories to be able to describe the observed four-dimensional word it is necessary for the extra dimensions to be compactified down to a size which we have not yet probed in particle experiments.    Theoretically, the cosmologically dimensional reduction [3,4] provide an interesting mechanism in which, during the cosmology evolution the three space we living are expanding while the extra spaces are contracting.   Whether  the mechanism can work or cannot is dependent of the matter of the universe and the topology of the spacetime  [3, 4].    According to the mechanism the extra spaces will be contracting to a small value, at which it is assumed that the quantum gravity effects could support it from collapsing into a singularity.    

    In this paper we have quantized the perfect-fluid cosmology in the 1+d+D dimensional Kaluza-Klein spacetimes by using the Schutz's variational formalism.   We have found the exact solutions of the Wheeler-Dewitt equation when the $d$ space is flat for the case with stiff fluid $p=\rho$, which can describe the matters in the early universe.   After the superposition of the solutions we study the time evolution of the scale factors in the later stage by computing the Bohmian trajectories of the ontological interpretation.  Our results show that the flat $d$ spaces and the compact $D$ spaces will eventually evolve into finite, non-zero scale functions.  However, the mechanism of cosmologically dimensional reduction does not shown in the stiff-matter spacetime.   We then turn to investigate the cases of  $\gamma \approx 2$.    We use the approximated wavefunction in the Wheeler-DeWitt equation to investigate the time evolution of the scale factors in the later stage.   Our results show that the flat $d$ spaces will be expanding forever and the scale function of the $D$ spaces is contracting which, however, does not become zero within finite time.  This means that the "crack-of-doom" singularity in which the extra dimensions goes to zero at finite time in the classical model [5] is now made to occur at $t=\infty$ in the quantum version investigated in this paper.

    Finally, we will mention that the investigations including more fundamental quantum fields besides the quantum perfect fluid [16] are more interesting and closing to the real world.   We hope to present the study in the near future.

%%%%%%%%%%%%%%%%%%%%%%%
\newpage
\begin{enumerate}
\item  M. J. Duff, B. E. W. Nilsson, and C. N. Pope, "Kaluza-Klein Supergravity", Phys. Rep. 130 (1986) 1.
\item M. B. Green, J. H. Schwarz, and  E. Witten,  {\it Superstring Theory}, Cambridge University Press, 1987;  J. Polchinski, {\it String Theory}, Cambridge University Press, 1998.
\item  A. Chods and S. Detweiler, Phys. Rev. D21 (1980) 2167;  P.G.O. Freund, Nucl. Phys. B 209 (1982) 146.
\item D. Bailin and A. Love,  Kaluza-Klein Theories, Rep. Prog. Phys, 50 (1987) 1087;  J. M. Overduin, P. S. Wesson, "Kaluza-Klein Gravity", Phys. Rep. 283 (1997) 303  [gr-qc/9805018].
\item  Sahdev, Phys. Letts. B 137 (1984) 155, Phys. Rev. D30 (1984) 2485. 
\item  V. G. Lapchinskii and V. A. Rubakov, Theor. Math. Phys. 33 (1977) 1076. 
\item N.  A. Lemos, " Radiation-Dominated Quantum Friedmann Models", J.Math.Phys. 37 (1996) 1449  [gr-qc/9511082 ];  F. G. Alvarenga, Nivaldo A. Lemos, "Dynamical Vacuum in Quantum Cosmology", Gen.Rel.Grav. 30 (1998) 681 [gr-qc/9802029].
\item J. A. de Barros,  N. Pinto-Neto, and M. A. Sagioro-Leal, "The Causal Interpretation of Dust and Radiation Fluids Non-Singular  Quantum Cosmologies"  Phys. Letts. A. 241 (1998) 229 [gr-qc/9710084]. 
\item F.G. Alvarenga, J.C. Fabris, N.A. Lemos, G.A. Monerat, "Quantum cosmological perfect fluid models", Class. Quantum Grav. 34 (2002) 651 [gr-qc/0106051].
\item F.G. Alvarenga, A.B. Batista, J.C. Fabris, S.V.B. Goncalves, "Troubles with quantum anisotropic cosmological models: Loss of unitarity ", Gen.Rel.Grav. 35 (2003) 1659-1677 [gr-qc/0304078]; "Anisotropic quantum cosmological models: a discrepancy between many-worlds and dBB interpretations" [gr-qc/0202009].
\item  B. F. Schutz, Phys. Rev. D2  (1970) 2762;  Phys. Rev. D4 (1971) 3559.
\item   D.  Brown, "Action functionals for relativistic perfect fluids" Class. Quant. Grav. 10 (1993) 1579 [gr-qc/9304026]; "On variational principle for gravitational perfect fluid", Annals Phys. 284  (1996) 1-33.
\item Y. B. Zeldorich, Sov. Phys. JETP 14 (1962) 1143;  J. B. Barrow, Nature 272 (1978) 211.
\item  C. W. Misner, K. S. Thorne, and J. A. Wheeler {\it Gravitation}
(San Fransisco: Freeman 1973)
\item  I. S. Gradshteyn and I. M. Ryzhik ,"Table of Intergals, Series and Products." Academic Press. New York 1980.
\item  C. Misner, "Minisuperspace"  in Magic without Magic: John Archibald Wheeler, (Freeman, 1972).
\item  N. A. Lemos and G. A. Monerat, "A quantum cosmological model with static and dynamical wormholes", Gen.Rel.Grav. 35 (2003) 423 [gr-qc/0210054].
\end{enumerate}
\end{document}